# Interfacial thermal conductance in graphene/black phosphorus heterogeneous structures

Yang Chen,[a] Yingyan Zhang,[c] Kun Cai,[a] Jinwu Jiang,[d] Jin-cheng Zheng,[e] Junhua Zhao,[b*] Ning Wei[a*]

a College of Water Resources and Architectural Engineering, Northwest A&F University, 712100 Yangling, China

b Jiangsu Key Laboratory of Advanced Food Manufacturing Equipment and Technology, Jiangnan University, 214122 Wuxi, China

c School of Computing, Engineering and Mathematics, Western Sydney University, Penrith, NSW 2751, Australia

d Shanghai Institute of Applied Mathematics and Mechanics, Shanghai Key Laboratory of Mechanics in Energy Engineering, Shanghai University, Shanghai

e Department of Physics, Xiamen University, 361005 Xiamen, China

Abstract

Graphene, as a passivation layer, can be used to protect the black phosphorus from the chemical reaction with surrounding oxygen and water. However, black phosphorus and graphene heterostructures have low efficiency of heat dissipation due to its intrinsic high thermal resistance at the interfaces. The accumulated energy from Joule heat has to be removed efficiently to avoid the malfunction of the devices. Therefore, it is of significance to investigate the interfacial thermal dissipation properties and manipulate the properties by interfacial engineering on demand. In this work, the interfacial thermal conductance between few-layer black phosphorus and graphene is studied extensively using molecular dynamics simulations. Two critical parameters, the critical power $P_{cr}$ to maintain thermal stability and the maximum heat power density $P_{max}$ with which the system can be loaded, are identified. Our results show that interfacial thermal conductance can be effectively tuned in a wide range with external strains and interracial defects. The compressive strain can enhance the interfacial thermal conductance by one order of magnitude, while interface defects give a two-fold increase. These findings could provide guidelines in heat dissipation and interfacial engineering for thermal conductance manipulation of black phosphorus-graphene heterostructure-based devices.

# 1. Introduction

Black phosphorous (BP) has stimulated tremendous research interest recently due to its outstanding electronic,[1-2] mechanical[3] and thermal properties.[4] Unlike graphene which suffers zero energy gap, few-layer BP possesses a direct energy gap in the electronic band. BP shows great advances on the high room-temperature transporting electron mobility, ~1000 $cm^2V^{-1}s^{-1}$ and large on/off current ratio in BP-based electronic device.[1-2] The intrinsic atomic structure of BP leads to interesting anisotropic mechanical and thermal properties as well as negative Poisson ratio.[3] Therefore, BP holds great promise for application of 2D semiconductor devices.

However, BP behaves chemically unstable in air.[5-6] In ambient environment the phosphorous atoms of BP can react with oxygen and water easily. To prevent such chemical reaction or isolate BP from oxygen and water, the most convenient and efficient method is wrapping/covering BP by other more environment-stable materials, such as graphene (Gr), boron nitride (BN) and dielectric/fluoropolymer.[7-8] In particular, graphene, which has high thermal conductivity, larger surface area and mechanical properties, is considered as the best candidate coverage.[9] For instance, the metal templates (such as Cu, Ni and Ru) can be protected from oxidation by depositing graphene on them.[10-11]

The long-term stability of BP covered by graphene in air exposure has been reported by Kim et al.[6] Their experimental results show that graphene can be used as a passivation layer to protect BP from chemical reaction in ambient environments. When BP is covered with few-layer graphene, thermal transport through the BP/Gr interface is the major route for heat dissipation. The BP/Gr interfaces is coupled by the weak van de Waals (vdWs) interactions suggests low thermal conductance due to the intrinsic interfacial thermal resistance at the interface.

Owing to thermal conductivity limits the maximum current density, the poor thermal performance of a heterogeneous structure limits its application in Micro/Nano-Electro-Mechanical Systems.[12] The accumulated Joule heat nucleates thermal hot spots, and potentially causes failure of devices. Therefore, to improve the performance of interfacial thermal conductance (ITC) becomes an urgent need.

Although thermal properties of BP, along both in-plane and cross-plane directions, have been well studied, their ITC with dissimilar materials, such as graphene have not been investigated systematically.

To improve the ITC, many approaches with various of techniques were developed by enhancing interfacial interaction or/and reducing the phonon spectra mismatch at the interface. For example, Gao and Müller-Plathe[13] reported that the ITC of graphene-polyamide-6,6 nanocomposites by surface-grafted polymer chains is improved remarkably and their ITCs increase continuously with the surface density of grafted chains.

Lin and Buehler[14] found that the alkyl-pyrene molecules, which possess phonon-spectra features of both graphene and octane, can act as phonon-spectra linkers to bridge the vibrational mismatch at the graphene/octane interface. The advantage of this approach is that the non-covalent functionalization does not induce additional defects or adatoms in graphene. It also enhances the thermal performance of nanocomposites while retaining the intrinsic mechanical and thermal properties of pristine graphene. Liu Ling and co-workers[15] reported that the thermal conductance across the interfaces between graphene and polymethyl methacrylate (PMMA) can be improved around 273% by introducing hydrogen-bond-capable hydroxyl groups to the interfaces. The coupling modes of the low-frequency vibration are enhanced. The improvement of the ITC is mainly attributed to the strengthening of the interfacial interactions. The energy between hydrogen bond interactions is around 1~2 orders of magnitude stronger than that of vdWs interactions. Furthermore, Ding et al.[16] reported that the thermal transport performance can also be effectively enhanced by introducing defects to interfaces. The reason is that the enhancement of interface friction improves the contribution of shear modes of phonon, which is closely related to the thermal conductance.

In this work, we studied ITC of a BP/Gr heterostructures using both non-equilibrium molecular dynamics (NEMD) and thermal relaxation methods. Each methods has its own advantages. The NEMD simulation provides steady state phonon transport information and is helpful in understanding the thermal conductivity of the overall heterostructures. The thermal relaxation method mimics experimental laser-based pump-probe process and is used to measure ITC. By using this approach, two important parameters are identified. One is the critical heat power density per area, $P_{cr}$, which could maintain the thermal stability of system, and the other is the maximum heat power density per area $P_{max}$, which can be loaded in the system. We also manipulate the interfacial thermal conductance by altering cross-plane compressive strain and interface defect engineering. The relationship between the interfacial thermal conductance and the phonon spectra at the interface is discussed in great details. Our work suggests that the interfacial thermal conductance can be effectively tuned in a wide range for different external strain and interfacial defects.

## 2. Modeling & Methodology

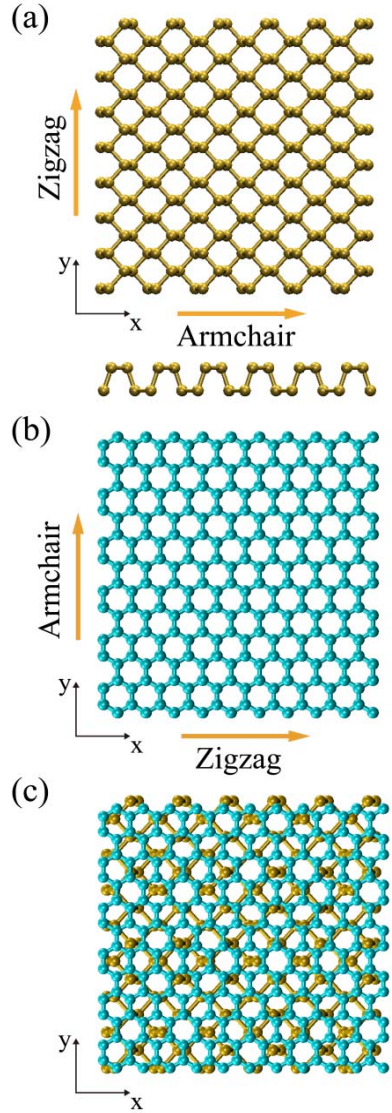

Figure 1. The schematic model of the BP/Gr interface between a graphene sheet and a single-layer of BP. (a) Atomic structure of a single-layer BP; (b) Atomic structure of a graphene sheet; (c) The combined structure of a single-layer BP and a graphene sheet.

To build and minimize the lattice mismatch of the heterostructure of BP and graphene, the supercell of BP sheet with 6 × 19 unit cells is chosen (with armchair edge along the *x*-direction, see Figure 1(a)). The graphene supercell is made of 15 × 11 unit cells (with zigzag edge along the *x*-direction see Figure 1(b)). Thus there exits small mismatch between BP and graphene in *x* and *y* directions (1.56% and 0.054%), respectively.

Based on the supercells of graphene and BP, BP/Gr heterostructures are created with the cross-section area about 2.67 nm × 6.30 nm (as shown in Figure 2(a, c)).

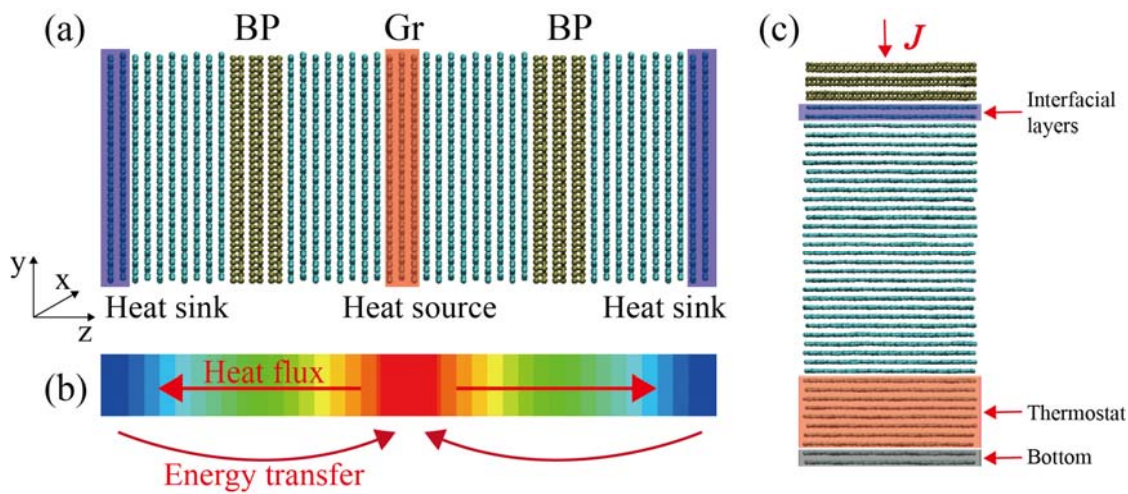

Figure 2. Schematic models for interfacial thermal conductance of graphene/BP heterostructures using periodic and free boundary condition models. (a) Using Müller-Plathe's (MP) method, the heat source and sink are located in the middle and at the two ends of the model under periodic boundary condition, respectively. (b) The kinetic energy exchanged between the heat source and the heat sink employing MP method is briefly described. (c) The MD model of graphene/BP heterostructures under free boundary condition using the thermal relaxation method.

All molecular dynamics (MD) simulations are performed using the large-scale atomic/molecular massively parallel simulator (LAMMPS) package.[17] The Stillinger-Weber (SW) potential[18] is employed to describe the covalent interaction between phosphorous atoms, while the interactions of carbon atoms in multilayer graphene are described by AIREBO potential.[19] The van der Waals interactions of interlayers are modeled with 12-6 Lennard-Jones (LJ) potential and the corresponding parameters are listed in Table I.[20] The cutoff distance of the LJ potential is set as 1.2 nm. A time step of 0.5 fs is employed in the MD simulations. Two simulation models, periodic (Figure 2(a)) and non-periodic (Figure 2(c)) boundary conditions in thermal transport directions, are employed to explore the ITC at BP/Gr interface using direct

NEMD and thermal relaxation methods, respectively.

Table 1
Values of parameters in the L-J potential for Carbon (C) and Phosphorous (P) atoms

| Atom 1 | Atom 2 | σ (Å)  | ε (eV)   |
|--------|--------|--------|----------|
| P      | C      | 3.4225 | 0.006878 |
| P      | P      | 3.438  | 0.015940 |
| C      | C      | 3.400  | 0.002844 |

2.1 NEMD-Periodic method

To calculate ITC based on the reverse NEMD or MP method,[21] the periodic boundary condition model (Figure 2(a)) is employed. In this model, periodic boundary conditions are applied in all the three dimensions. The initial configuration is firstly equilibrated at NPT (constant temperature and pressure) with room temperature at 300K and pressure at 1 bar for 0.5 ns ($1\times10^6$ times steps). Afterwards, the system is then switched to be at the microcanonical ensemble (NVE). The ITC between BP and graphene layers is investigated by MP approach. The system is equally divided into 40 slabs along the heat transport direction (z-dimension), with the heat source and heat sink selected in the middle and at the ends of the model, respectively, as shown in Figure 2(a, b). The heat flux is added by exchanging the kinetic energies between the hottest atom in the heat sink slab and the coldest atom in the heat source slab, the value of which can be obtained by:

$$J = \frac{\sum_{Nswap} \frac{1}{2}(mv_h^2 - mv_c^2)}{t_{swap}}, \quad (1)$$

where $N_{swap}$ and $t_{swap}$ are the entire swap time and number of swaps, while, $v_h$ and $v_c$ are the atomic velocities of the hottest and coldest atoms, respectively. To ensure that the system can reach the non-equilibrium steady state before extracting, we monitor the temperature of each layer with time (supporting info. Figure S1.). The temperature $T$ refers to the average temperature of all atoms in each layer

$$T_i(slab) = \frac{2}{3Nk_B} \sum_j \frac{p_j^2}{2m}. \quad (2)$$

As shown in supporting information Figure. S1, the system reaches a non-equilibrium state after 2.0 ns. The temperature distribution is obtained by calculating the average data over the following 3 ns, as shown in Figure 3 (a). The thermal conductance $G$ is calculated using $G = J/(2 \times \Delta T \times A)$, where $J$ is the heat flux, $A$ is the cross-section area and $\Delta T$ is the temperature difference at the interface and the factor 2 is due to the heat current propagating in two directions. We illustrated the time-averaged temperature profile in Figure S2 (see Supporting info.), it shows that there is greater temperature jump $\Delta T$ at the interface close to the heat sink. It demonstrates a smaller thermal conductance at this interface. Overall, this result is in good agreement with the previous experimental and theoretical studies.[22-24] Here, we take the value $\Delta T$ by averaging two interface temperature jumps $\Delta T = (\Delta T_{hot} + \Delta T_{cold})/2$.

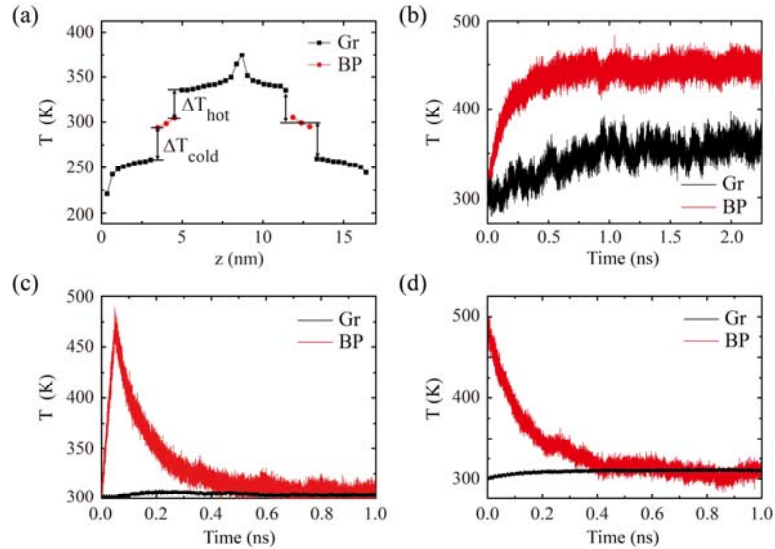

Figure 3. (a) The temperature profile of BP/Gr heterostructure, which is obtained after the system reaches non-equilibrium steady state via MP method. (b) Profile of temperature evolution in BP sheets (red line) and graphene buffer layer (black line) with heat flux continuous injected into BP layers. (c, d) Profile of temperature evolution in BP sheets (red line) and graphene buffer layer (black line) using thermal relaxation method with (c) pump-probe approach and (d) BP layers rescaled at initial higher temperature method.

## 2.2 NEMD-Non-Periodic method

Another direct NEMD simulation is conducted using the non-periodic boundary condition model (Figure 2(c)), in which the periodic boundary conditions are applied along the in-plane (x and y) directions, with the free-boundary condition along the out-of-plane (z) direction. Three layers of BP are covered on multilayer graphene sheets (including 40 sheets of graphene around 15 nm in height) through vdWs interactions. The top two graphene sheets neighboring BP layer are defined as buffer layer, and the motion of the bottom layer of graphene is in restrained in the out-of-plane direction. The temperature at a 2 nm thick slab nearby the bottom layer maintains 300 K using Berendsen thermostat as a heat sink in the simulations.[12]

To generate the temperature difference $\Delta T$ at Grapheneand BP interface, a continuous heat flux $J$ is injected into BP layers by adjusting the velocity of phosphorus atoms according to[25]:

$$v'_i = v_T + \alpha(v_i - v_T), \qquad (3)$$

where $v'_i$ and $v_i$ are the current and initial velocities of thermostat atoms, respectively, in time iteration. $v_T$ is the velocity of the center of mass of the thermostat, and $\alpha$ is the rescaling factor, which yeilds

$$\alpha = \sqrt{1 \pm \frac{\Delta\varepsilon}{E_R}}, \qquad (4)$$

where $\Delta\varepsilon$ is the injected/subtracted energy from specified atoms and $E_R$ is the relative kinetic energy

$$E_R = \sum_i \frac{1}{2} m_i (v_i^2 - v_T^2), \qquad (5)$$

and the heat flux $J = \Delta\varepsilon / (A \Delta t)$. Based on the continuous heating by the supported BP sheets. The temperature of BP layers increases firstly and then achieves an equilibrium plateau temperature within tens of picosecond by dissipating the energy across the interface into the graphene layers via the interface through vdWs interactions. After achieving equilibrium plateau temperature, the ITC can be obtained by using: $G = J/\Delta T$, where $\Delta T$ is the averaged temperature difference at BP/Gr interfaces nearby the hot and cold thermostats.

2.3 Thermal relaxation method

In the thermal relaxation studies, the non-periodic boundary condition model (Figure 2(c)) is employed. The initial configuration is first equilibrated at 300K by performing constant volume and constant temperature (NVT) for 1.0 ns ($2\times10^6$ times steps). Afterwards, the system is then switched to a NVE ensemble. The initial temperature gap can be generated by imposing a short heat pulse with area power density $P$ in the range from 0.25 to 0.6 GWm$^{-2}$ on BP layers for a few picoseconds, or increasing the temperature of BP layers instantaneously to a specified value by rescaling the velocities of phosphorus atoms.[25] After the heat source is being removed, the temperature in BP layer decays exponentially within a few hundred picoseconds, as shown in Figure 3(c, d), respectively.

The relaxation time $\tau$ is obtained by exponential fitting ($\Delta T(t) = \Delta T(t_0)\exp[(t_0 - t)/\tau]$) the temperature history of BP. The value of ITC is calculated via $G = C/(A\times\tau)$, where $C$ is the heat capacity of BP. The heat capacity variation of BP with temperature is presented in the Supporting info. (see Figure S3). The simulation of thermal relaxation approach mimics the experimental pump-probe process and is enough to track the thermal dissipation. Besides, we perform a trial simulation with larger model comprising 80 layers of graphene sheets. It is found that the model has similar thermal conductance with merely 3% different in light of this fact, we use the smaller model with 40-layer graphene in following simulations for the sake of computation efficiency. To get the substantial insight into interfacial energy transfer mechanism, we investigate the phonon vibration spectra of the atoms in the graphene and BP layers. The vibrational density of states (VDOS) $P(\omega)$ at the frequency $\omega$ which can be calculated by performing the Fourier transform on the velocity auto-correlation function as[26]

$$P(\omega) = \frac{1}{\sqrt{2\pi}} \int_0^\infty e^{i\omega t} \left\langle \sum_{j=1}^N v_j(t) v_j(0) \right\rangle d\omega \qquad (6)$$

To quantify the match between the vibration spectra of atoms at the interfaces, an overlap factor $S$ is defined based on the correlation parameter,[27] which is employed to explore the insight of interfacial interactions.

$$S = \frac{\int_0^\infty P_{Gr}(\omega) P_{BP}(\omega) d\omega}{\int_0^\infty P_{Gr}(\omega) d\omega \int_0^\infty P_{BP}(\omega) d\omega} \qquad (7)$$

Where, $P_{Gr}(\omega)$ and $P_{BP}(\omega)$ denotes the phonon spectra at frequency $\omega$ of graphene and BP atoms at the interface. The overlap factor S indicates the degree of match in the phonon spectra.

**3 Results and discussions:**
3.1 Interfacial thermal conductance
The properties of ITC at BP/Gr interfaces are investigated using NEMD and thermal relaxation methods and the results are plotted in Figure 4. For the NEMD simulation with MP method, the heat flux imposed on the system is adjusted by changing the kinetic energy exchange frequency (in the range of 100-1000 time step) and then the relationship between ITC and $\Delta T$ can be obtained. The temperature difference $\Delta T$ at the interface has negligible effect on the ITC value, which is almost invariable and their average value is about 50.23 $MWm^{-2}K^{-1}$. This phenomenon is consistent with previous studies.[22, 28] Whereas, the obtained ITC from another NEMD simulation using continuous heating method shows almost linearly increasing with $\Delta T$ (the slop is 0.059 by linear fitting). The various $\Delta T$ at the interface is generated by changing the injected power areal density $P$. Such phenomenon has also been reported for the graphene on SiC substrate.[12] The distinct tendencies of ITC obtained from above two methods are caused as follows: Despite the increasing $\Delta T$, the temperature of BP layers is almost invariable and keeps a constant temperature of 300K using MP method. The temperature of BP layers increasing with $\Delta T$ using continuous heating method, which may cause the temperature effect on the interfacial thermal transport. We will discuss this effect in details in the following section. Figure 4(b) shows the studies of ITC using thermal relaxation method by heating a short time or setting a fixed temperature. One can see that the ITC obtained from the thermal relaxation method is about 22 MWm$^{-2}$K$^{-1}$, which is independent of the initial heating or setting temperature. The magnitude of ITC obtained by the thermal relaxation method is smaller than that from NMED, which agrees well with the previous studies.[12] The lower value of ITC is due to the fact that the low frequency phonons make the majority contribution in the ITC, while the energy is adding to the high frequency modes mostly in the thermal relaxation method.[29]

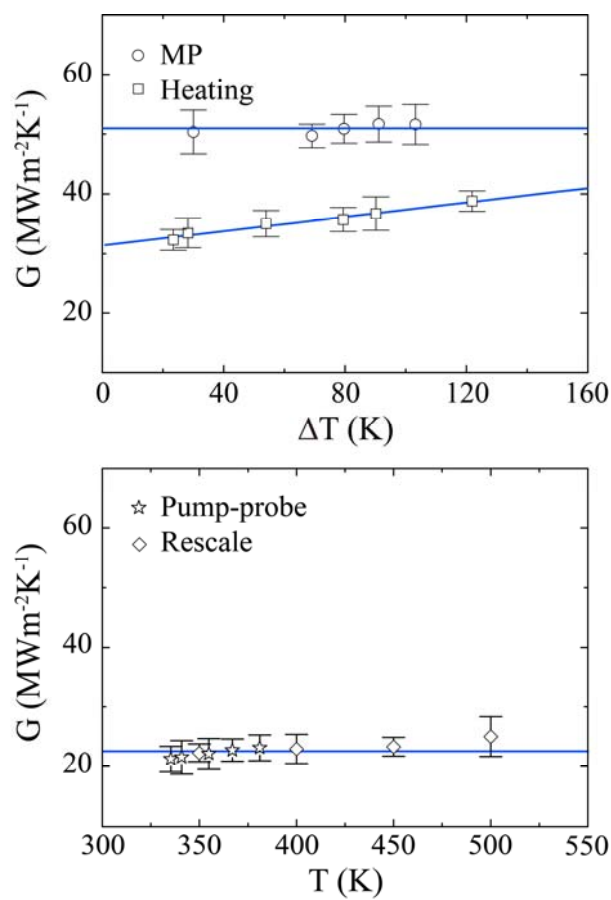

Figure 4. The thermal conductance of BP/Gr interface obtained using various methods. (a) NEMD simulations with MP method and the relationship between ITC and temperature jump $\Delta T$ adjusted by changing the exchange frequency of kinetic energy between heat source and heat sink (labeled as "MP"). NEMD simulations by adding different non-translational kinetic energy $P$ to BP layer (labeled as "Heating"). (b) The interfacial thermal conductance as a function of temperature of BP before relaxation

performed. The thermal relaxation simulations with heat-relaxation (labeled as "Pump-probe") and temperature rescale-relaxation methods (labeled as "Rescale"), respectively.

3.2 Heat dissipation process

To explore the relationship between inject power areal density $P$ and interfacial temperature different $\Delta T$ in the NMED with non-periodic model method, the time evolution of local temperature in both BP and graphene buffer layer are plotted in Figure 5(a-c). Results show that $P$ cannot induce explicit temperature difference (or obvious temperature jump at the interface) when $P$ is small enough, as shown in Figure 5(a). Here, we define the maximum value of $P$ which cannot induce obvious temperature jump at the interface as $P_{cr}$. When $P > P_{cr}$, both $T_{BP}$ and $T_{Gr}$ first rise up within a few nanosecond and then the system reaches equilibrium steady state (see Figure 5(b, c)). Subsequently, a temperature difference $\Delta T$ will form at the interface and it increases with $P$ monotonically. In order to get the exact value of $P_{cr}$ which is important to the practical application of BP device, we extract temperature of BP after the system reaches equilibrium with various $P$ and the results are presented in Figure 5(d). The relationship between $T_{BP}$ and $P$ can be fitted well by parabolic function (see Fig.S4 in Supporting info.).[12] However, when $P$ is lower than the critical value, $P_{cr}$, the $T$-$P$ relationship deviates the parabolic function, as shown in the insect of Figure 5(d). Therefore, we obtain the value of $P_{cr}$ is 0.28 GW/m$^2$. Another limit value of $P$ is the maximum load of the system before the device failure. The melting temperature of BP obtained from our MD simulation is about 520K. According to the $T$-$P$ relationship, the corresponding $P$ is 6.52 GW/m$^2$. We then define this value as the maximum areal power density the system can be loaded, labeled as $P_{max}$, as shown in Figure 5(d).

To obtain physical insights from the simulation results, we derive the linear relationship of thermal resistance $R$ and the areal power density $P$ (the details of derivation can be found in Supporting info.), as presented in Figure 6, $R = 1/G = c_1P+c_2$. We can also obtain the relationship between thermal conductance $G$ and interfacial temperature difference $\Delta T$ as follows (details for the extraction of the above two equations are given in supporting):

$$G = \frac{2}{c_2 + \sqrt{c_2^2 + 4c_1\Delta T}} \quad (8)$$

Therefore, a fixed inject power density $P$ has a certain corresponding thermal resistance/conductance. Here, the values of interfacial thermal conductance $G$ at $\Delta T = 0$ (obtained according to their linear relationship in Figure 4) are employed due to they are sensitive to $\Delta T$. Figure 6 shows the thermal conductance/resistance along with $\Delta T$. We can see that the predictions are well consistent with the MD results. To verify the linear fitting used in NEMD simulation results in Figure 4(a), we study their relationship of $P$ in the range of from 0.2 to 6.0 GW/m² (our present MD studies region). We find that their relationship can be roughly described by a linear relationship and the values of $G$ at $\Delta T = 0$ obtained by linear extension and equation 8. have only less than 5% different.

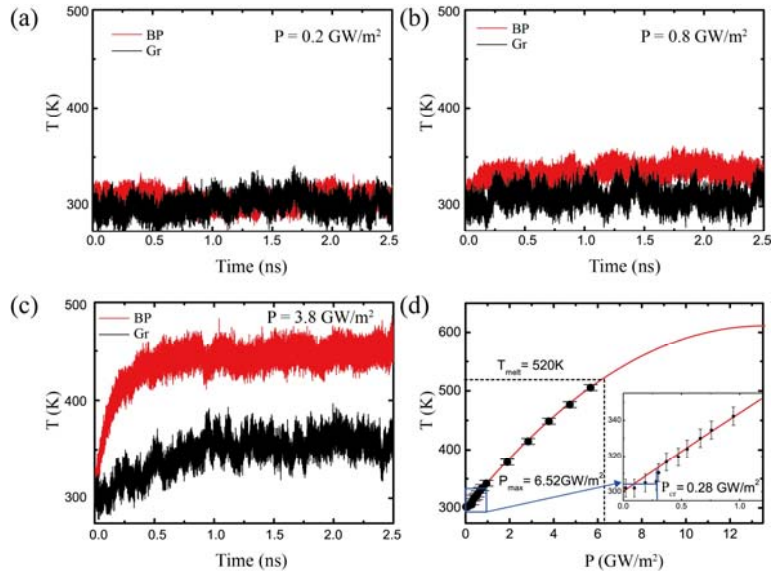

Figure 5. (a-c) Time evolution of temperature in BP and graphene using heating method at various power density $P$. The steady state of thermalization is established in 1.5 ns. (d) The temperature of BP as a function of $P$ and the critical power $P_{cr} = 0.28$ GW/m² to maintain thermal stability and the maximum load power $P_{max} = 6.52$ GW/m² of BP are predicted.

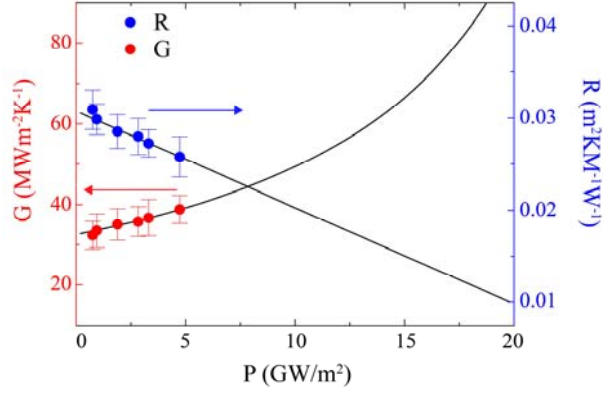

Figure 6. Thermal conductance *G* and thermal resistance *R* as a function of generated power density *P*.

3.3 Size effect

Due to the advantages of MP method in the interfacial thermal transport properties studies, independent of $\Delta T$, convenient for study of interfacial engineering effect and time-saving in simulations, the following MD simulations are performed using MP method. We then examine the system size effect on the prediction of BP/Gr interfacial thermal conduction. We find that ITC is insensitive to the cross-section size, the value of *G* obtained from twofold cross-section dimension model has less than 5% different. Then we study the dependence of ITC on the thickness of block graphene and BP, respectively. Figure 7 shows that magnitude of ITC increases with the increasing of number of graphene layers $N_{Gr}$ with the number of BP layers $N_{BP}$ fixed at 3. The increasing of ITC with $N_{Gr}$ is attributed to the phonon mean free path in graphite along the cross-plane direction being greater than the thickness of bulk graphene layers. The enhancement of ITC with the thickness of bulk graphite has also been reported at solid/solid and solid/liquid interfaces.[22, 30] Our results show that the ITC increases with $N_{Gr}$ and this relation can be described by an exponential function, and the maximum $G = 89.08$ MWm$^{-2}$K$^{-1}$ is obtained at $N_{Gr} = \infty$ according to this exponential relationship.

Then we study the size effect with increasing $N_{BP}$ with fixed $N_{Gr}$ at 20. The simulation results show that ITC is not sensitive to the thickness of BP layers. The values of ITC with 3-layer and 11-layer BP are 50.87 MWm$^{-2}$K$^{-1}$ and 50.07 MWm$^{-2}$K$^{-1}$, respectively. The negligible size effect of BP on ITC suggests that the phonon mean free path of block BP along the cross-plane direction is short. In our following MP simulation model, $N_{BP}$ = 3 and $N_{Gr}$ = 20 are used to study the ITC at BP/Gr interface.

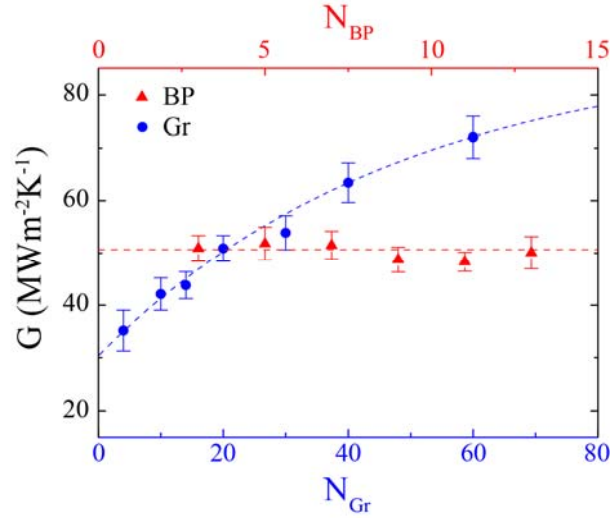

Figure 7 Interfacial thermal conductance $G$ at BP/Gr interface at room temperature versus number of BP layers (red triangle, with graphene layers $N_{Gr}$ = 20) and graphene layers (blue dot, with BP layers $N_{BP}$ = 3), respectively. The relation between $G$ and $N_{Gr}$ can be fitted by an exponent function $G = ae^{-N_{Gr}/c} + b$ ($a$ = -58.73, $b$ = 89.08, $c$ = 48.34).

3.4. Temperature effects

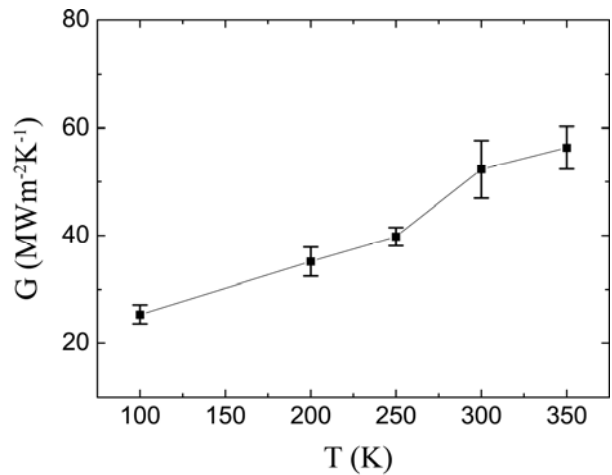

Figure 8. Interfacial thermal conductance $G$ as a function of environmental

temperature *T*.

We further explore the dependence of *G* on the environmental temperature *T* (see Figure 8). It shows that the value of *G* increases with temperature *T*, from 25.33 MWm$^{-2}$K$^{-1}$ at 100K to 56.34 MWm$^{-2}$K$^{-1}$ at 350K. It features an almost linear dependence relationship for the Umklapp processes which play the critical role in the heat transport.[31-32] This tendency is in good agreement with those for interfaces between graphene and SiO$_2$ or polymer materials. To get the substantial insight and a detailed understanding of the heat transfer mechanisms at the interface, we investigate the vibration spectra density of states at interfaces by analyzing the correlations of atomic vibrations. Figure 9 shows that the out-of-plane phonon spectra of graphene and BP make a greater contribution to the overall interfacial thermal transport and the overlap vibration spectra mainly in the low frequency zone, range from 1 to 15 THz. We also calculate the overlap parameter *S*, which related to ITC, using equation 7. It is interesting to notice that fewer phonons are excited in both graphene and BP at lower temperature (200K), which induce a lower value of *S* (0.304) and which limits heat transport. When *T* grows, more phonons, especially the low frequency phonon spectra of BP, are excited and involved into the interfacial thermal transport. The power spectra and the correlation factors at various temperature backgrounds (200K, 300K and 400K) are shown in Figure 9.

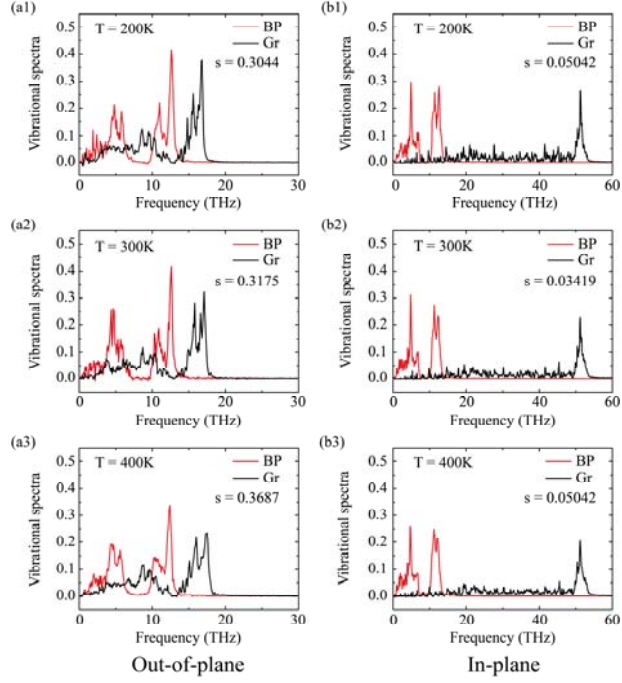

Figure 9 Out-of-plane and In-plane vibrational spectra of BP and graphene atoms at interface under different background temperature.

3.5 Effect of cross-plane compressive strain

Our present MD studies predict a poor inferior thermal performance at BP/Gr interface (in the range of 21.22- 89.08 $MWm^{-2}K^{-1}$), which is similar to several other graphene based interfacial systems, e,g., 50 $MWm^{-2}K^{-1}$ for the graphene-water system,[22] 14 $MWm^{-2}K^{-1}$ for the graphene-$MoS_2$ system,[16] 21 $MWm^{-2}K^{-1}$ for graphene-resin system,[29] and 43.23 $MWm^{-2}K^{-1}$ for graphene-SiC system.[12] They are much smaller than that of graphite layer thermal conductance by two orders of magnitude. Therefore, it's important to enhance the ITC for practical applications. Applying a cross-plane strain is an effective method to tune the heat transport across interface performance. It has been proved in previous studies of graphene,[30] $MoS_2$[16] and BP [4] *etc.* two dimensional systems. To study the strain effect on cross-plane heat transport at BP/Gr interface, we apply uniaxial cross-plane compressive strain to the periodic boundary condition model (Figure 2(a)). The engineering strain is defined as: $\varepsilon = (l-l_0)/l_0$, where $l_0$ and $l$ are the initial and finial length of the box at the cross-plane direction, respectively. The stress-strain relationship in heterostructure is illustrated in Figure S6 (see supporting info.). The cross-plane stress increases monotonically with compressive strain $\varepsilon$ until the system fails under compression at $\varepsilon_{max}$ = 18.65%. The system failure at BP layers due to the weaker stability of BP structure (see Figure S6 inset). Then we studied the strain effect on ITC using MP method. The variations of thermal conductance at room temperature under various cross-plane strains are plotted in Figure 10. The MD results of $G$ are normalized by the strain-free value. In present work, we only illustrate the MD results of strain from 0.0% (unstrained) to -9.1% because the BP layers are unstable under large strain and require short integral time step and longer relaxation time in NEMD simulations.

The compressive strain is found to enhance thermal conductance $G$ and follows an exponential dependence relationship, which is consistent with previous studies.[16, 30] As the compressive strain is $\varepsilon_{max}$ = 18.65%, we can predict the maximum enhancement factor of $G$ is 16.13 according to this exponential relationship. The enhancement of $G$ under compressive strain can be attributed to two main reasons. One is the compression on the multi-layer structures reduces the interlayer distance, and therefore increases the interlayer interactions, i.e., enhances the LJ interactions and coupling between interlayers and leads to the increasing the heat transport performance.[4, 30, 32] The other is that increasing shear interaction between interlayers will increase the thermal transport contribution from in-plane phonons.

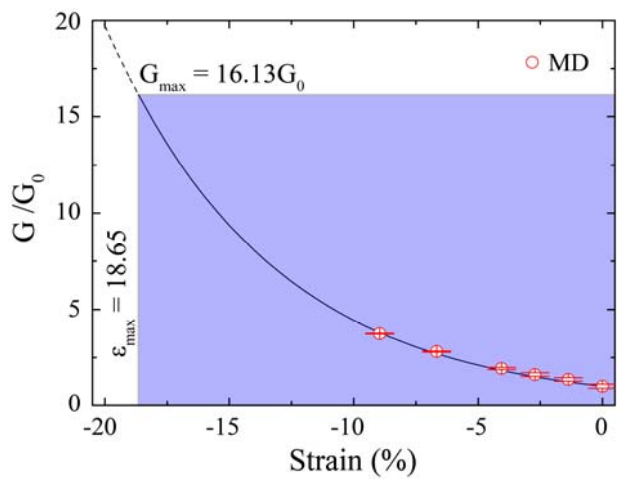

Figure 10. Varying range of the thermal conductance along with cross-plane compressive strain.

The power spectra of carbon and phorphorous atoms in graphene and BP layers at the interface under various strains are shown in Figure 11. Interestingly, we find that the phonon spectra of graphene at both in-plane and out-of-plane components remain almost unchanged, while the peak of out-of-plane phonon density of state (DOS) of BP exhibits a blue shift, e.g., from 12.3 THz at $\varepsilon = 0.0$ to 14.4 THz at $\varepsilon = -11.93\%$. The reason is that the LJ interactions between BP layers is weaker than that of between graphene layers or between graphene and BP layers. A similar effect on phonon spectra shift induced by strain has been reported by previous studies.[16, 30, 33-34] The compression results in stiffening the phonons transferring across interlayers, i.e., both phonon group velocity and specific heat increase, and lead to an enhancement of interlayer thermal conductance. The variation of overlap parameter $S$ with compressive strain is plotted in Figure 12. We can see that phonon coupling enhanced at both in-plane and out-of-plane components, and $S$ increases with the performed strain. The findings are well consistent with our conclusions mentioned above. Hence, the ITC can be effectively modulated by cross-plane strain.

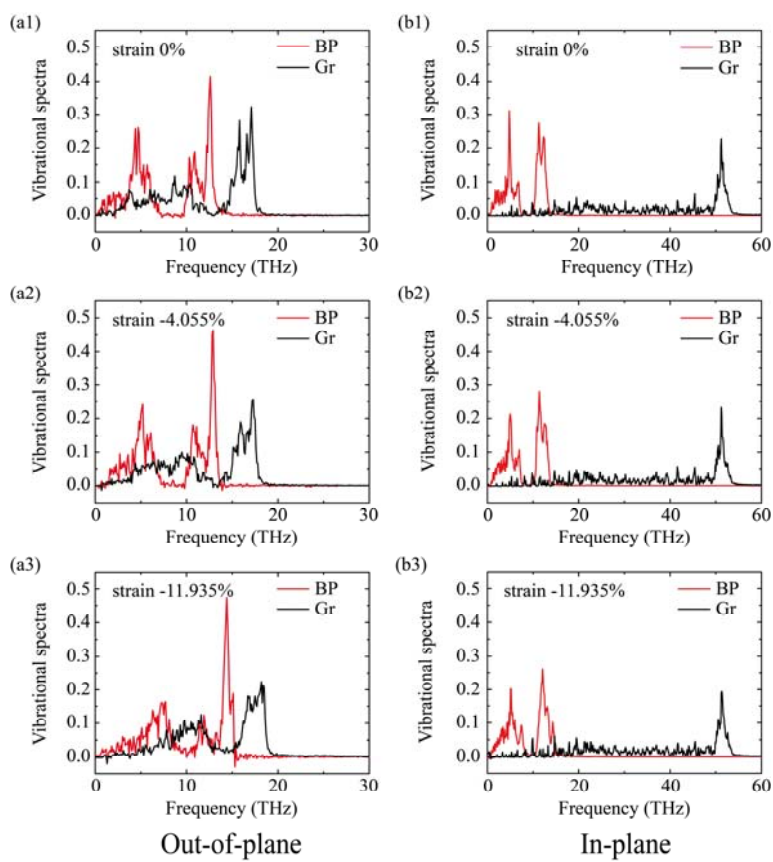

Figure 11. Vibration density of states of BP under various cross-plane strains.

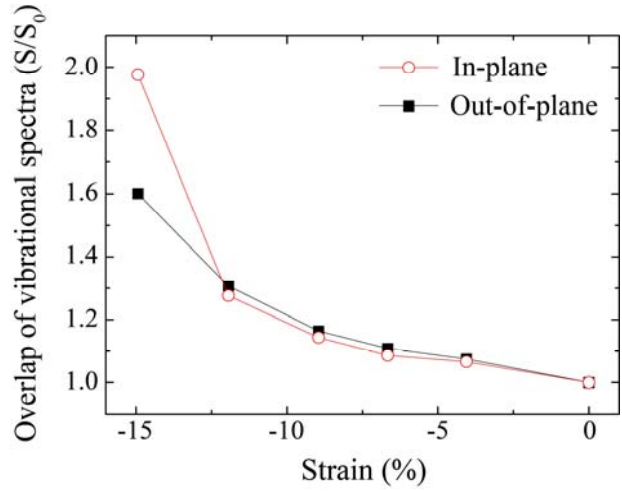

Figure 12. Normalized overlap parameter $S/S_0$ of vibration spectra at interfaces along with cross-plane compressive strain. $S_0$ are the overlap parameter of strain-free values which are 0.034 and 0.317 for in-plane and out-of-plane, respectively.

3.6 Defects effect

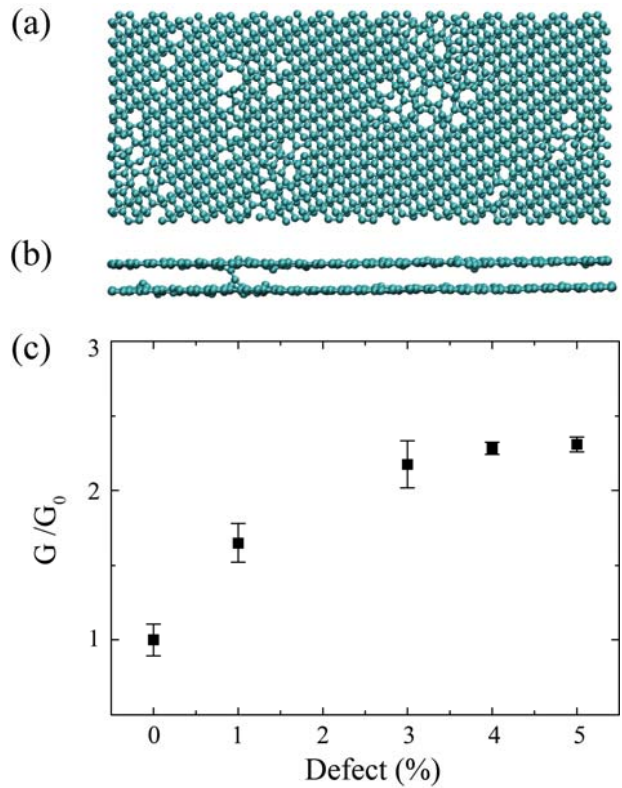

Figure 13. Enhancement of thermal conductance by interface defective concentration. (a, b) The top-view and side-view of surface defective at 5%. The defects are set by removing carbon atoms on two graphene layers neighboring BP layers. (c) Thermal conductance as a function of interface defective concentration.

In our previous studies, defect engineering at graphene and MoS$_2$ interface presents promising opportunities to increase the heat transfer performance across the interfaces, As vacancy defect concentration in graphene reaches 5% the ITC increases by 180%.[16] Here, we study the effect of interface vacancy defective concentration on ITC by removing carbon atoms of two graphene layers nearby BP layers. The top and side views of graphene with vacancy defect of 5% are presented in Figure 13 (a, b). The MD results of $G$ are normalized by the defect-free value $G_0$ plotted in Figure 13(c). It indicates that $G$ increases with defective concentration monotonically. Defect free graphene is a good lubricant material due to its ultra-low friction coefficient with both liquid and solid interfaces. The superlubricity properties of graphene have been reported in both experimental and theoretical studies.[35] According to our previous studies, due to the low friction between layers at interface, the transmission coefficient of shearing modes is slight. Therefore, introducing vacancy defects in graphene layer at interface increasing the roughness and friction between graphene and BP layers and leading to the increasing the transmission of shear modes of phonons. Figure 14 illustrates the surface potential distributions with various vacancy defect concentrations in graphene sheets. The potential profile at the first layer position of BP sheet along with the black line labeled in Figure 14(a1-a3) are plotted in Figure 14(c1-c3).

By introducing vacancy defects to the graphene layers at interface, the lattice symmetry and incommensurate between graphene and BP layers are broken. The magnitude of surface potential barrier $\Delta E = E_{max} - E_{min}$ of defective graphene surface is one order larger than that of defect-free graphene. The enhancement of friction at the interface improves the transmission coefficient of shear modes, and further leads to the growth of overall interfacial thermal transport performance. Defects, therefore, are major reason for the increasing of contribution of in-plane phonons to thermal conductance. This conclusion can also be verified by the overlap vibration spectra parameters $S$ shown in Figure 15. However, the out-of-plane phonon DOS seems insensitive to vacancy defects (see Figure S7). As the defects in interfacial graphene layers increase the shear interactions, and the coupling of low-frequency vibrational modes are enhanced as shown in Figure S7. Figure 15 shows that *S* of in-plane component increases with defect concentration density and it is about threefold at defect density concentration 5% than that of the defect free one, while value of *S* of out-of-plane has an enhancement by 125% only. Therefore, our results show that interfacial defective engineering is an effective approach to enhance the ITC.

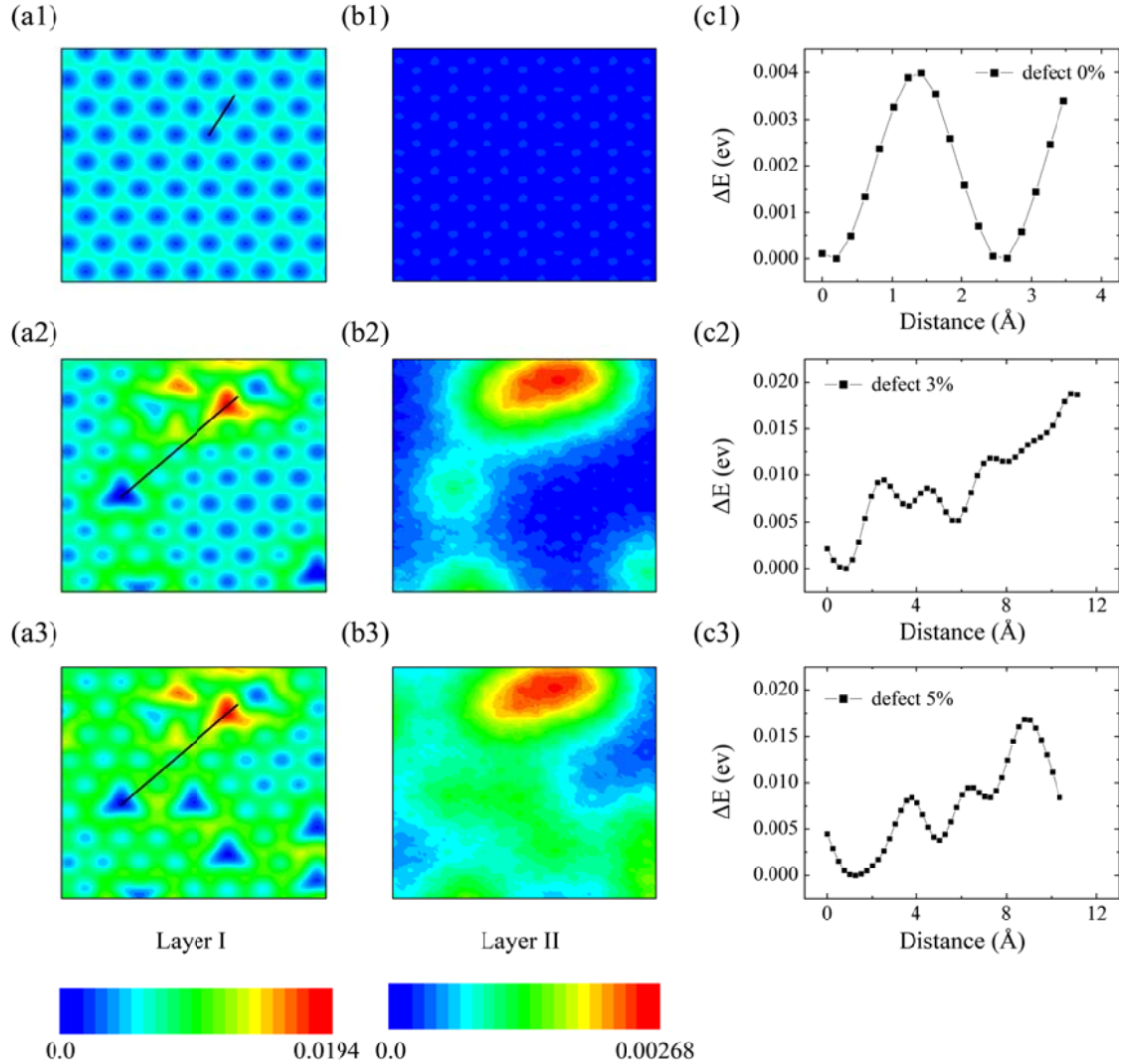

Figure 14. Potential distribution at the height of the first (a1-a3) and second (b1-b3) phosphorus atomic layer of BP sheet covered on graphene substrate. (c1-c3) Energy barrier profiles on the first phosphorus atomic layer along with the labeled black line in (a1-a3).

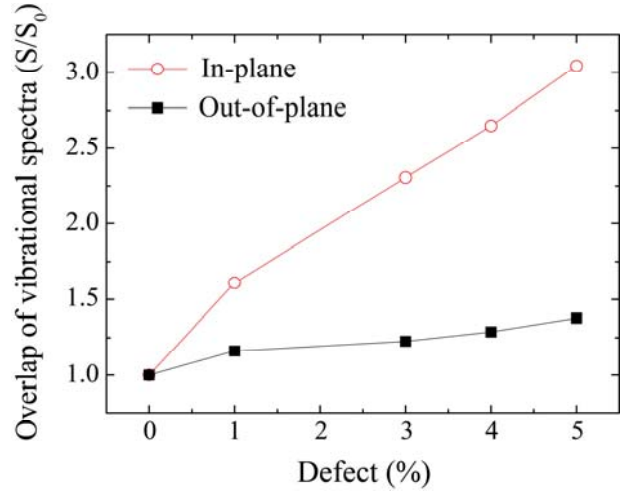

Figure 15. Normalized overlap of vibration spectra $S/S_0$ as a function of interfacial defective concentration.

## 4. Conclusions

Interfacial thermal conductance plays a key role in composite structure due to the drastically compromising overall thermal dissipation properties. In this work, we investigate the thermal properties across BP/Gr interface using atomic molecular dynamics simulations. The critical power $P_{cr}$ to maintain thermal stability and the maximum load power $P_{max}$ of BP are identified. It is demonstrated that the ITC can be tuned in a wide range by introducing strain or defects in interface. The improved ITC by cross-plane strain and defects can be attributed to the increasing of overlap of interfacial phonon vibration spectra. Under the cross-plane strain, the value of overlap phonon spectra in out-of-plane and in-plane components increasing with almost the same tendency. While, in-plane phonon overlap magnitude increases much faster than that of out-of-plane component for the case of interfacial defects. Our findings suggest effective methods for drastically improving interfacial thermal conductance towards wide applications.

Acknowledgements

We gratefully acknowledge support by the National Natural Science Foundation of China (Grant No.11572140, 11302084, 11502217, U1332105, 11204253, and 11335006 ), the Programs of Innovation and Entrepreneurship of Jiangsu Province, the Fundamental Research Funds for the Central Universities (Grant No. JUSRP11529), Postdoctoral Science Foundation(No. 2015M570854 and 2016T90949 ) Open Fund of Key Laboratory for Intelligent Nano Materials and Devices of the Ministry of Education (NUAA) (Grant No. INMD-2015M01) and "Thousand Youth Talents Plan". W. N. thanks Prof. Zheyong Fan in Aalto University for helpful discussion on vibrational density of states.

Supporting Information

# Interfacial thermal conductance in graphene/black phosphorus heterogeneous structures

Yang Chen,[a] Yingyan Zhang,[c] Kun Cai,[a] Jinwu Jiang,[d] Jin-cheng Zheng,[e] Junhua Zhao,[b*] Ning Wei[a*]

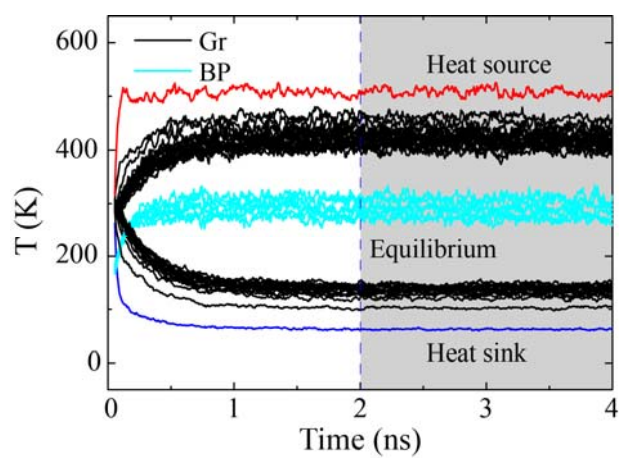

Figure S1. Time evolution of temperature (T) of each graphene and BP layer using MP method.

It should be noted that the temperature jump at a high temperature interface (closer to the heat source) is smaller than that the low temperature one (closer to the heat sink), as shown in Figure S2. It is consistent with previous study.[21] Here, we deal with $\Delta T$ the by averaging the two temperature jumps interfaces $\Delta T = (\Delta T_{hot} + \Delta T_{cold})/2$.

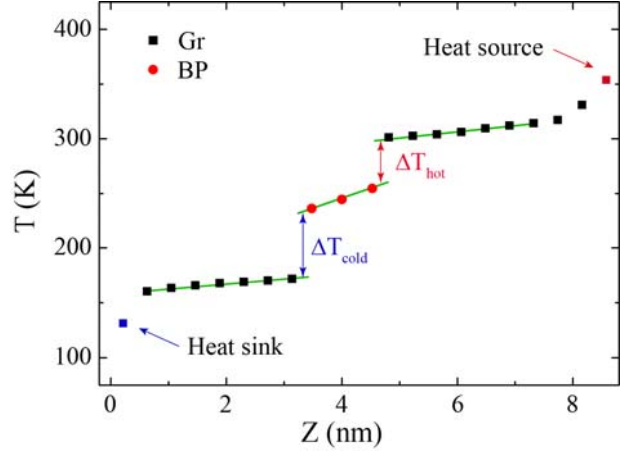

Figure S2. The temperature profile of the system using MP method. Two temperature jumps $\Delta T_{\text{hot}}$ closer to the heat source and $\Delta T_{\text{cold}}$ closer to the heat sink, corresponding to interfaces between BP and graphene are presented.

We use GULP to calculate the vibrational partition function,

$$Z_{vib} = \sum_k w_k \sum_\tau \frac{1}{1-e^{-\frac{\hbar\omega_\tau}{k_B T}}}, \quad (S1)$$

where the first summation is over all $k$ points with the weight $w_k$. The second summation is taken over all phonon modes.

The vibrational partition function can be used to compute the heat capacity at constant volume,

$$C_v = RT\left[2\left(\frac{\partial LnZ_{vib}}{\partial T}\right) + T\left(\frac{\partial^2 LnZ_{vib}}{\partial T^2}\right)\right], \quad (S2)$$

where $R$ is the gas constant, and the results are presented in Figure A3. The value obtained by above function is 21.04 J/mol/K at 300K while it is 26.61 J/mol/K from MD simulation. Here, we adopt value of 26.61 J/mol/K in our calculation of interfacial thermal conductance using thermal relaxation method.

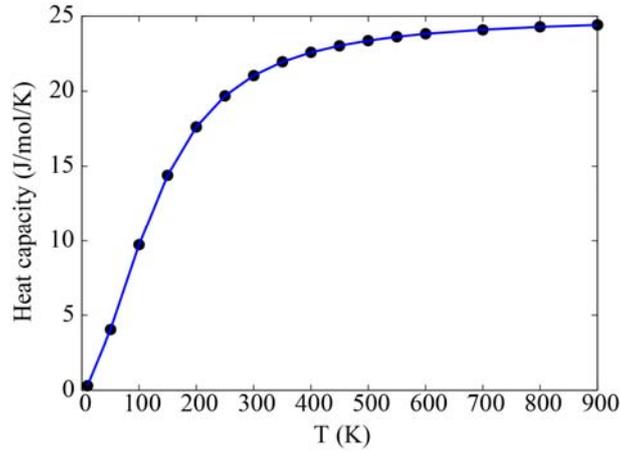

Figure S3. Heat capacity of BP as a function of temperature.

The equilibrium temperature of BP and graphene using heating method can be well fitted by parabolic functions, see Figure S4. Thereby, we can get the relationship of thermal conductance $G$ (resistance $R$) and power density P according to the following derivations.

$$T_{BP} = a_1 P^2 + a_2 P + a_3 \tag{S3}$$

$$T_{Gr} = b_1 P^2 + b_2 P + b_3 \quad (a_3 = b_3 = 300) \tag{S4}$$

$$\Delta T = T_{BP} - T_{Gr} = c_1 P^2 + c_2 P \; (c_1 = a_1-b_1,\; c_2 = a_2-b_2) \tag{S5}$$

and the thermal resistance $R = \Delta T/J$, here $J = P$ when the system reaches equilibrium. Then the linear relationship between thermal resistance $R$ and $P$ can be presented as: $R = \Delta T/P = c_1 P + c_2 = 1/G$, see Figure 5. In addition, the thermal conductance $G$ with $\Delta T$ can be written as:

$$G = \frac{2}{c_2 + \sqrt{c_2^2 + 4c_1 \Delta T}} \tag{S6}$$

see Figure S5. We can see that in our studied region the curve is almost linear and only a little different for the extrapolation value at $\Delta T = 0$ (they are 32.52 MWm$^{-2}$K$^{-1}$ and 31.41 MWm$^{-2}$K$^{-1}$ from equation S4 and linear fitting, respectively).

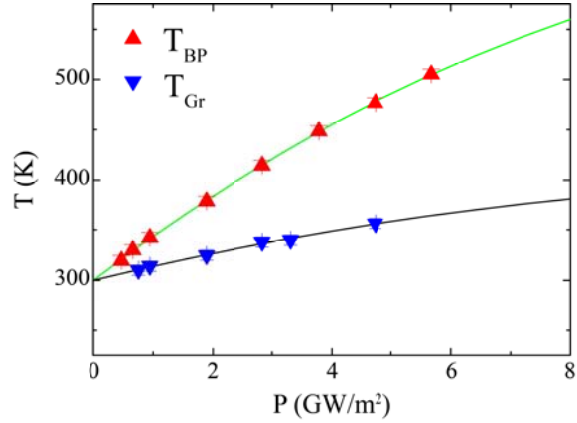

Figure S4. Temperature of BP and graphene (buffer layer) as a function of *P*.

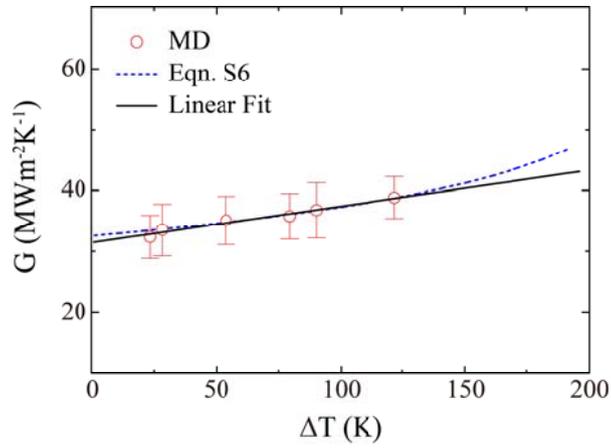

Figure S5. Thermal conductance as a function of temperature different Δ*T* at the interface.

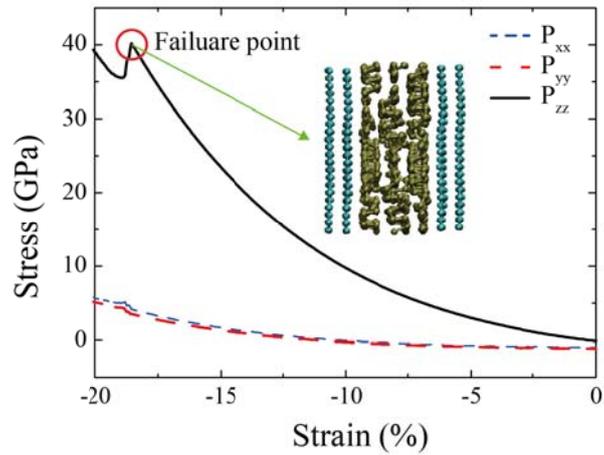

Figure S6. Stress-strain relation versus uniaxial cross-plane strain in BP and graphene heterostructure.

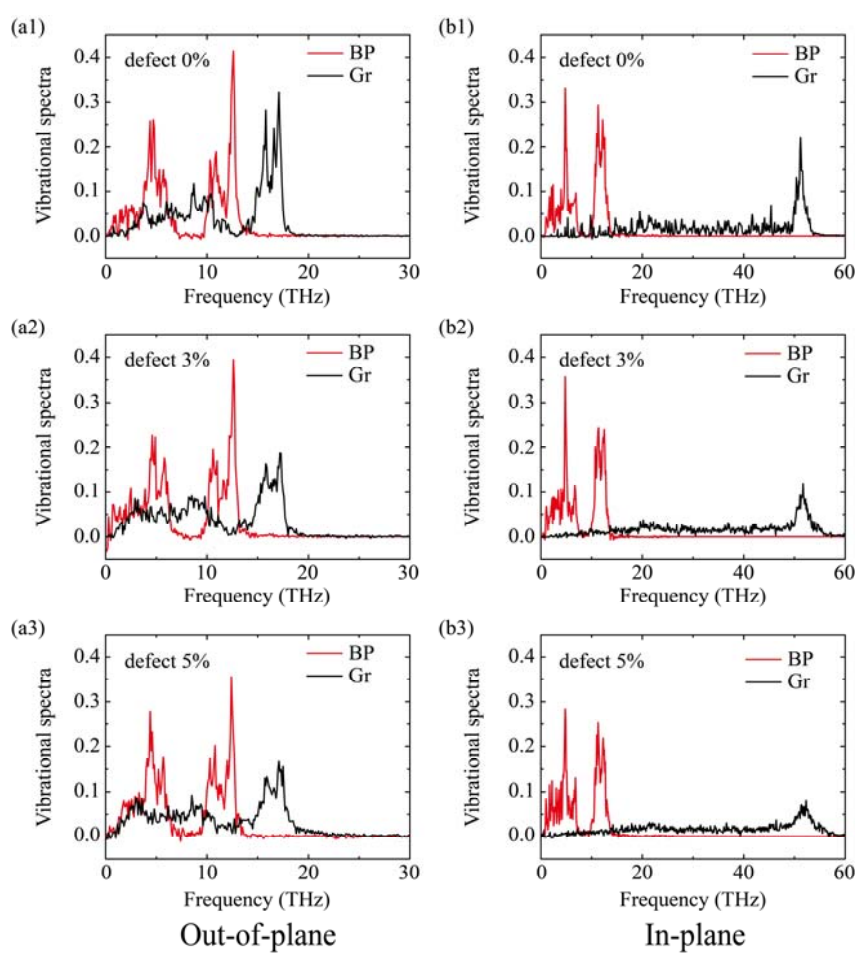

Figure S7. Vibration density of states of BP with various defect concentration.